\documentclass[a4paper,11pt]{article}
\usepackage{jinstpub} 
\usepackage{lineno}

\usepackage[acronym]{glossaries}
\setacronymstyle{long-short}
\newacronym{awg}{AWG}{arbitrary waveform generator}
\newacronym{jj}{JJ}{Josephson junction}
\newacronym{sfq}{SFQ}{single flux quantum}
\newacronym{snspd}{SNSPD}{superconducting nanowire single photon detector}
\newacronym{dsp}{DSP}{digital signal processing}
\newacronym{pxi}{PXI}{Peripheral Component Interconnect eXtensions for Instrumentation}
\newacronym{cots}{COTS}{commercial-off-the-shelf}
\newacronym{qick}{QICK}{Quantum Instrumentation Control Kit}
\newacronym{rfsoc}{RFSoC}{Radio-Frequency System-on-Chip}
\newacronym{rfdac}{RFDAC}{radio-frequency digital-to-analog converter}
\newacronym{rfadc}{RFADC}{radio-frequency analog-to-digital converter}
\newacronym{pl}{PL}{programmable logic}
\newacronym{ps}{PS}{processing system}
\newacronym{dma}{DMA}{direct memory access}
\newacronym{dds}{DDS}{direct digital synthesis}
\newacronym{cci}{CCI}{cache coherent interconnect (CCI)}
\newacronym{fpga}{FPGA}{field-programmable gate array}
\newacronym{sfdr}{SFDR}{spur-free dynamic range}
\newacronym{sinad}{SINAD}{signal-to-noise and distortion}
\newacronym{vga}{VGA}{variable-gain amplifier}
\newacronym{pcb}{PCB}{printed circuit board}
\newacronym{rsfq}{RSFQ}{rapid single flux quantum}
\newacronym{aqfp}{AQFP}{adiabatic quantum flux parametron}
\makenoidxglossaries

\usepackage{layouts}
\title{\boldmath Time-tagging data acquisition system for testing superconducting electronics based on an RFSoC and custom analog frontend}

\author[1]{R. A. Foster,\note{Corresponding author.}}
\author{S. Kandeh,}
\author{O. Medeiros,}
\author{A. Simon,}
\author{M. Castellani,}
\author{and K. K. Berggren}
\affiliation{Massachusetts Institute of Technology,\\
77 Massachusetts Ave, Cambridge MA, USA}

\emailAdd{reedf@mit.edu}

\abstract{

Novel electronic devices can often be operated in a plethora of ways, which makes testing circuits comprised of them difficult.
Often, no single tool can simultaneously analyze the operating margins, maximum speed, and failure modes of a circuit, particularly when the intended behavior of subcomponents of the circuit is not standardized.
This work demonstrates a cost-effective time-domain data acquisition system for electronic circuits that enables more intricate verification techniques than are practical with conventional experimental setups.
We use high-speed digital-to-analog converters and real-time multi-gigasample-per-second waveform processing to push experimental circuits beyond their maximum operating speed.
Our custom time-tagging data capture firmware reduces memory requirements and can be used to determine when errors occur.
The firmware is combined with a thermal-noise-limited analog frontend with \ensuremath{{50}\,\mathrm{dB}} of dynamic range.
Compared to currently available commercial test equipment that is seven times more expensive, this data acquisition system was able to operate a superconducting shift register at a nearly three-times-higher clock frequency (\ensuremath{{200}\,\mathrm{MHz}} vs. \ensuremath{{80}\,\mathrm{MHz}}).
}

\keywords{Digital electronic circuits, Digital signal processing (DSP), Data acquisition circuits, Data acquisition concepts, Data reduction methods}

\arxivnumber{2505.21714} 

\newcommand{\SI}[2]{\ensuremath{{#1}\,\mathrm{#2}}}

\begin{document}

\maketitle
\flushbottom

\section{Introduction}
\label{sec:intro}

Improving the speed and energy efficiency of integrated circuits is an active area of research, inviting approaches that range from device-level improvements \cite{wachter_microprocessor_2017, peng_carbon_2019}, to modification at the device- and circuit-level \cite{polonsky_new_1993, ayala_mana_2021}, to entirely new processing architectures that leverage unique capabilities of novel devices \cite{li_analogue_2018}.
With the exception of \cite{li_analogue_2018}, which used a custom-built data acquisition system for testing, the full circuit in each of these works was operated well below its maximum operating speed, and only a single device or simplified version of the circuit was tested up to a higher speed.
In these specific cases, as is the case for many works involving novel electronic devices, limitations of test equipment can prevent testing of full circuit operation at high speeds.
However, this limitation raises a question of whether the complex circuit can actually work at the speed of the simpler circuit, or if emergent properties of these highly-integrated circuits limit their speed.

Standardized interfaces and logic levels of the experimental circuit allow researchers to leverage specialized high-speed test equipment, like bit-error-rate testers, which helps address this concern.
For example, a low-power superconducting microprocessor based on \glspl{aqfp} was demonstrated at gigahertz clock speeds \cite{ayala_mana_2021}.
However, even in this case, limitations of the programmability of the test equipment meant that experimenters could only test a single input and single output of their circuit at a time.
While this provided reassurance that their circuit could operate at high speeds in these specific cases, they unfortunately could not test the behavior of their circuit while simultaneously toggling multiple inputs.
Building on the approach of standardized interfaces, once a technology is somewhat mature, one can construct circuits that test themselves. In fact, the first experimental demonstrations of \gls{rsfq} circuits used DC-to-single-flux-quantum (DC-to-SFQ) converters to generate and process the inputs and outputs of the circuit under test \cite{koshelets_experimental_1987,kaplunenko_experimental_1989}.
However, one must be completely sure that the testing circuit is working, otherwise its output cannot be trusted.
In the case of DC-to-SFQ converters, sufficient theoretical understanding of Josephson junctions and the simplicity of the circuit makes it reasonable to assume it works correctly.
Once these basic self-tests are proven to work, more complex testing circuits can progressively be constructed \cite{polonsky_new_1993}.
To address some of these challenges faced by researchers developing circuits with novel devices that may behave in new or unexpected ways, we introduce a cost-effective, high-performance, integrated data acquisition system for testing electronic circuits, particularly ones with multiple parallel inputs and outputs.

To demonstrate the effectiveness of our data acquisition system, we use it to test a binary shift register made of superconducting nanocryotrons \cite{mccaughan_superconducting-nanowire_2014}.
Circuits based on nanocryotrons could improve the capabilities of superconducting photon and particle detectors \cite{steinhauer_progress_2021, polakovic_unconventional_2020}, however, the complexity and scale of nanocryotron circuits demonstrated to date has been limited to relatively small circuits, with the exception of an encoder circuit constructed from more than 100 nanocryotrons \cite{huang_monolithic_2024}.
The encoder is a promising demonstration for nanocryotron technology, and the testing apparatus presented in this work will further accelerate the development of more complex detector readout schemes and edge processing circuits.


With the exception of a few cases \cite{butters_scalable_2021, foster_superconducting_2023, buzzi_nanocryotron_2023, castellani_nanocryotron_2024}, all previously demonstrated nanocryotron circuits, including the encoder, were tested with a multichannel oscilloscope and one or more \glspl{awg} \cite{mccaughan_superconducting-nanowire_2014,zhao_compact_2018,zheng_superconducting_2020,toomey_superconducting_2020,huang_splitter_2023,huang_monolithic_2024,wang_attojoule_2025,castellani_superconducting_2025}.
This approach works well for circuits with purely combinational inputs and outputs.
However, even with the sophisticated triggering mechanisms of modern digital storage oscilloscopes, testing failure modes of stateful logic can quickly become impossible without saving and post-processing many tens to hundreds of gigabytes of waveform data.
This is undesirable because more than 90\% of the time in an experiment is spent moving data instead of actually measuring the circuit under test.\footnote{This 90\% figure assumes that acquisition occurs at \SI{1}{GS/s} with a bit depth of \SI{10}{b} and that data transfer is able to saturate a gigabit ethernet connection, which is an optimistic assumption. In practice this may be more like 99\% or higher depending on required sampling rate and the ability of the oscilloscope to saturate the network connection.}
This reduces the scope of operating margin analyses that can be performed in a reasonable amount of time.
Furthermore, such a large fraction of time moving data instead of measuring the circuit leads to blind time where errors in the circuit may occur without detection.
Integrated solutions that combine signal generation and capture into a single piece of equipment mitigate slow data movement by keeping data in fast random access memory or even processing it in real time.
Such an approach to acquisition allows for faster and more complex measurements, such as bit error rate as a
function of applied magnetic field, temperature, and bias current.

Commercial test equipment based on the \gls{pxi} bus protocol can provide tighter integration than separate \gls{awg}s and oscilloscopes \cite{keysight_pxi_nodate, national_instruments_ni_nodate}, which can increase testing speed by several orders of magnitude.
For example, a \gls{cots} \gls{pxi} digitizer and \gls{awg} from Keysight were used to test a binary shift register made with superconducting nanocryotrons, reducing the amount of time to perform a bit-error-rate measurement by 50,000$\times$ compared to the conventional setup \cite{foster_superconducting_2023}.
Shift registers are key building blocks for digital readout of pixelated arrays, serving either as in-pixel ring-buffers or performing serialization of pixel data.
Therefore, understanding the limits of operating speed of shift registers is critical for engineering the highest-performance digital readout schemes for imagers.
However, sub-{GS/s} sample rates of the \gls{cots} equipment used in \cite{foster_superconducting_2023} prevented the shift register from reaching its intrinsic maximum operating frequency.
Unfortunately, there is not a cost-effective \gls{cots} solution between the sub-{GS/s} and ultra-wideband \ensuremath{{20}\,\mathrm{GS/s}} offerings at the present time.

A number of custom systems have been designed specifically for testing classical and quantum superconducting electronics.
The Octopux is designed for testing \gls{jj}-based circuits, and as a result, its inability to drive high-impedance loads \cite{zinoviev_octopux_1997} limits its usefulness for testing nanocryotron circuits, where experimental devices can exhibit resistances exceeding \ensuremath{{1}\,\mathrm{k\Omega}} when driven out of the superconducting state.
Nanocryotron memory testers have been developed \cite{butters_scalable_2021,butters_digital_2022}, but overly-specialized firmware limits their applicability to circuits beyond cryogenic memory.
Several data acquisition systems based on the \gls{rfsoc} have been demonstrated for frequency-multiplexed readout of quantum and classical superconducting circuits, such as qubits and microwave photon detectors \cite{stefanazzi_qick_2022, smith_mkidgen3_2024, liu_development_2024}.
These solutions leverage the extremely wideband \glspl{rfadc} and \glspl{rfdac} of the \gls{rfsoc} to control and readout thousands of devices on a single wire. 
Although it is possible to bypass the downconversion step in some of these systems to perform time-domain acquisition at baseband, their firmware would need to be modified substantially to capture data at the full sample rate of the \glspl{rfadc}, which is required for fast nanocryotron devices that can generate pulses shorter than \SI{1}{ns} \cite{kerman_kinetic_inductance_2006}.
Another system based on an \gls{rfsoc} has demonstrated full-rate time-domain readout of \glspl{snspd} with custom firmware \cite{xie_entangled_2023}.
Due to the lack of readily-available firmware for \glspl{rfsoc} that fits the requirements for testing nanocryotrons, we chose to develop a custom data acquisition system based on the \gls{rfsoc} platform due to its low-noise, high sample rate data converters and extensive programmable logic resources.

By leveraging the advantages of the RFSoC platform, our data acquisition system can not only assist bringup and troubleshooting of nanocryotron circuits, but also push them beyond their maximum operating speed.
This work uses the ZCU111 development board with the XCZU28DR, a 1$^{\rm st}$-gen \gls{rfsoc} with eight 14-bit \ensuremath{{6}\,\mathrm{GS/s}} \glspl{rfdac} and eight 12-bit \ensuremath{{4}\,\mathrm{GS/s}} \glspl{rfadc}.
A custom analog frontend interfaces the \glspl{rfdac} and \glspl{rfadc} with our experimental circuits.
The data converters are integrated on the SoC and rely on a stripped-down AXI-stream interface for transferring data to and from the \gls{fpga} fabric (programmable logic) \cite{pg269}.
The software and firmware for this project, as well as the schematics for the analog frontend are released under an MIT license \cite{foster_rfsoc_daq}.
The firmware includes a configurable, multichannel level discriminator and timetagger to allow it to only save sample data for events of interest.
This reduces further waveform processing and memory requirements while still preserving snippets of the original, high-resolution waveforms that can give researchers key insight into the operation of their circuit.
Furthermore, the high sample rate capability of the signal generation and acquisition paths allows complex testing procedures to be performed on experimental circuits, pushing them to their limits under realistic stimuli.
This allows one to verify that the ensemble of devices in a circuit can work together at a high clock rate, rather than verifying a complex circuit at a few kHz clock speed and extrapolating its performance based on measured speeds of a single device or simpler circuit.

\section{Data acquisition system architecture}



The \gls{rfsoc} on the ZCU111 development board contains two parts: a multi-core ARM processor --- the processing system, and a \gls{fpga} --- the programmable logic.
The programmable logic interfaces with the \glspl{rfdac} and \glspl{rfadc} which convert between digital and analog signals.
Custom firmware configures the programmable logic to perform real-time processing of digitized waveforms.
A custom analog frontend interfaces the differential analog signaling used by the RF dataconverters with conventional single-ended cabling used for interfacing with experimental superconducting devices.




\begin{figure}[htbp]
\centering
\includegraphics[]{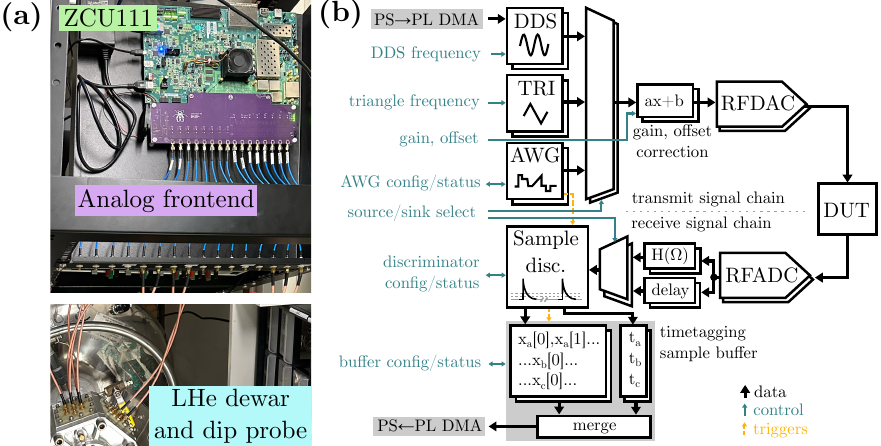}
\caption{
Diagram of data acquisition system.
\textbf{(a)}  Birds-eye-view of experimental setup with liquid helium dewar. Custom firmware and software runs on a ZCU111 \acrfull{rfsoc} development board, which is connected to a dip probe \cite{butters_digital_2022} and experimental circuit through a custom analog frontend.
\textbf{(b)} Block diagram of programmable logic as configured for this application.
To communicate with the ARM processor, \acrfull{dma} is used for bulk transfers of data; status and configuration registers are interfaced with individual reads and writes to short address ranges.
The data source for the \acrfull{rfdac} is selectable between a \acrfull{dds} sinewave generator, a triangle wave generator (TRI), and \acrfull{awg}.
The output of the \gls{rfdac} then passes to the device under test (DUT).
The signal from the DUT digitized by the \acrfull{rfadc} can optionally be passed through a filter or delay before it is sent to the sample discriminator, which is used to only save events of interest and discard other samples.
The output of the sample discriminator is saved in a buffer that can be transferred to larger off-chip memory via \gls{dma}.
\label{fig:arch}
}
\end{figure}

The custom data acquisition system is shown in Figure \ref{fig:arch}a, connected to a dip probe \cite{butters_digital_2022} housing a superconducting circuit in a liquid helium dewar.
A low-noise amplifier is included on each \gls{rfadc} channel of the analog frontend to reduce the requirements for external components when constructing experimental setups.
More detailed discussion of the analog frontend is included in Section \ref{sec:afe}.
A block diagram of the firmware and how it interfaces with the software running on the multicore ARM processing system is shown in Figure \ref{fig:arch}b.
The configuration and status of the various modules (e.g. gain/offset correction, sample discriminator, sample buffer, etc.) in the firmware are memory mapped, allowing them to be updated and monitored by the processing system (e.g. changing the gain/offset correction on one of the \gls{rfdac} channels, adjusting a threshold for the sample discriminator, or checking the number of samples saved in the buffer).
High-speed bulk transfers of data are performed with \gls{dma}, which uses a dedicated address range separate from the status and configuration registers.


\subsection{Software}

Software that runs on the ARM processor is used to orchestrate experiments, controlling the various modules in the programmable logic.
All of the software is written in Python and leverages PYNQ, a framework for Xilinx SoCs that enables development with Python instead of the standard C/C\texttt{++} workflow that is typically used.
The processing system runs a custom Linux image provided by PYNQ that includes the necessary libraries and packages for loading the firmware onto the programmable logic and interfacing with the memory-mapped modules described in the firmware.
PYNQ Linux is also configured with a daemon that spawns a Jupyter/IPython kernel at startup which can be connected to via local network for remote control of experiments.
See \cite{noauthor_pynq_nodate} and chapter 22 of \cite{crockett_exploring_2019} for more information on the PYNQ framework.

\subsection{Field-programmable gate array (FPGA) firmware}

Custom \gls{fpga} firmware configures the programmable logic to generate samples for the \glspl{rfdac} and processes the \gls{rfadc} data in realtime.
Figure \ref{fig:arch}b shows the organization of the firmware.
The firmware is broken into two halves: a transmit signal chain that generates signals to supply to the \glspl{rfdac}, and a receive signal chain that processes and stores signals from the \glspl{rfadc}.
Each individual module in these signal chains has configuration and status registers that can be written to and queried by the processing system through memory-mapped I/O.
The status and configuration registers are detailed in \cite{foster_rfsoc_daq}.

The firmware is written in an extensible way, making it easy to introduce additional modules into both the receive and transmit signal chains to perform data processing.
Because of the real-time nature of data converters, the signal paths do not apply backpressure with a ready signal: only a data and valid signal are necessary.
A ready signal is unnecessary because the data converters are guaranteed to produce or consume samples at a fixed rate; if one tries to stall the output of the \gls{rfadc}, there is no way to recover any samples it lost.
This design choice simplifies timing restrictions for deep pipelines, which can limit performance and complicate design when a full handshake with a ready signal is required \cite{abbas_latency_2018}.

\subsection{Signal generation}

In order to generate a stimulus for testing superconducting electronics, an \gls{awg} was implemented.
The \gls{awg} stores waveform data in on-chip block RAM, a cascadable SRAM technology used in Xilinx \gls{fpga}s \cite{mehta_xilinx_nodate}, and uses a simple state machine for handling \gls{dma} transfers of software-generated waveform data into the waveform buffers.
Each channel can store waveforms up to \ensuremath{{32}\,\mathrm{kS}} in length, or about \ensuremath{{5}\,\mathrm{\upmu s}} at \ensuremath{{6.144}\,\mathrm{GS/s}}.
The storage capacity could be increased by leveraging other memory technologies available, such as UltraRAM \cite{wp477}.
Channel memory can be cyclically read out up to $2^{64}-1$ times, and the cycle timing can be coordinated across multiple channels if desired.
The \gls{awg} can be configured at runtime to output a trigger signal to the sample buffer to initiate capture when it starts outputting data to the \glspl{rfdac}, which provides deterministic timing for measuring a device's response.

In addition to the \gls{awg}, triangle wave and \gls{dds} generators are also available to generate test waveforms.
The \gls{dds} was implemented with phase dithering to spread out spurs from phase quantization \cite{foster_scaling_2023}, achieving over \ensuremath{{100}\,\mathrm{dB}} \gls{sfdr} with a 4096-entry lookup table.
The high \gls{sfdr} was important for characterization of the analog frontend performance to ensure that any spurious tones and noise measured by the \gls{rfadc} came from the analog frontend and were not already present in the test signal.

\subsection{Signal acquisition}
\label{sec:sigacq}

As shown in Figure \ref{fig:arch}b, the data from the \glspl{rfadc} passes through optional filtering, a sample discriminator to only save events of interest, and a sample buffer.
The sample discriminator performs a hysteretic threshold comparison of the incoming waveform data: once the incoming data exceeds an initial threshold, the data is saved until it drops below a second threshold.
The sample buffer stores timing information of the events of interest (when the incoming data exceeds the initial threshold), as well as the data itself.

The sample buffer is implemented with the same on-chip block RAM as the \gls{awg} waveform buffers.
The choice to use SRAM instead of off-chip DRAM was done in order to allow the user to record data simultaneously from all eight \glspl{rfadc} running at full rate.
This requires a peak write bandwidth of \ensuremath{{393}\,\mathrm{Gb/s}}; more than twice the bandwidth of all off-chip DRAM interfaces on the ZCU111 board combined.
The current design allows data to be saved continuously for \ensuremath{{16}\,\mathrm{\upmu s}} (up to \ensuremath{{128}\,\mathrm{\upmu s}} when saving data from only one channel).
However, the memory capacity could be increased at least four-fold (\ensuremath{{64}\,\mathrm{\upmu s}} for eight channels or \ensuremath{{0.5}\,\mathrm{ms}} for one channel): the current design uses about 11\% of the available block RAM resources for the sample buffer and 27\% for other modules, leaving 62\% of it free.
This was done to limit the potential impact of routing congestion due to high utilization, however the practical limit for block RAM utilization in this architecture was not studied carefully.
An additional factor of two in capacity could be gained by using the other on-chip UltraRAM resource.\footnote{UltraRAM is a synchronous memory \cite{wp477}, and the sample buffers are implemented as true dual-port asynchronous memories to decouple the read/write bandwidth. However, by wrapping an UltraRAM with a clock-domain-crossing interface on the read port, it could be used in the sample buffer with no changes to the sample buffer logic.}

Given the limited capacity of on-chip memory, the sample discriminator is a critical element because it limits the amount of data that must be saved and post-processed after an experiment is done.
Eventually, the trace data must be analyzed to determine if an event of interest (e.g. a switching event in a superconducting nanowire) occurred.
One might then consider storing the minimum amount of information: whether or not an event occurred (and perhaps the time at which it occurred), which would dramatically reduce the memory capacity and bandwidth requirements for data acquisition.
However, saving the raw voltage data from the \gls{rfadc} can be very useful when one is debugging a circuit that is not working.
For example, if a superconducting circuit incorrectly switches once in a million times, it would be very useful to have the actual voltage traces associated with that error saved instead of hoping to reproduce the error in a subsequent measurement in which one saved all of the waveform data.
Furthermore, saving only the error prevents one from having to sift through a large amount of data to find the error.
\begin{figure}[htbp]
\includegraphics[]{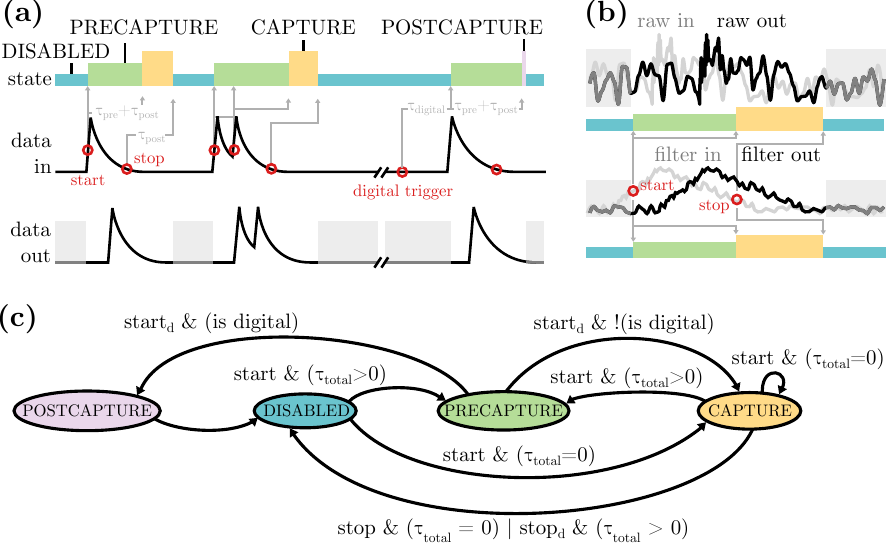}
\caption{
Operating principle of realtime sample discriminator for data reduction.
\textbf{(a)} Input data is compared against thresholds, and the current state is tracked to allow for hysteresis.
The output data is delayed to allow samples to be saved before the start event.
The valid signal is pulled low when the discriminator is in the DISABLED state, as indicated by the gray shaded regions on the ``data out'' waveform.
\textbf{(b)} Sharing of start/stop events between multiple channels is possible by changing the event source for the state machine of each channel.
In this example, we wish to save raw \gls{rfadc} data only when the output of a matched filter is above some threshold.
\textbf{(c)} Diagram of state machine transition rules. The state machine is initialized into the DISABLED state. When in any state other than DISABLED, the output valid signal is high, allowing the output data to be saved. The states and criteria for transitions are described in more detail in Section \ref{sec:discfsm}.
\label{fig:disc}
}
\end{figure}

\begin{figure}[htbp]
\includegraphics[]{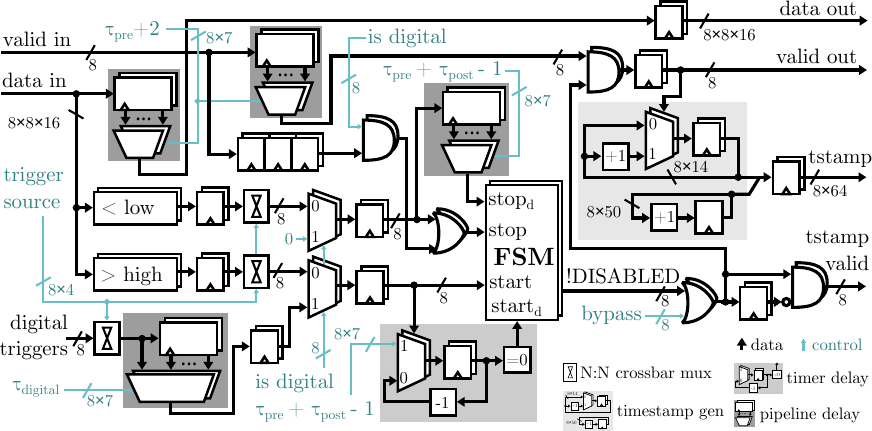}
\caption{Dataflow diagram of realtime sample discriminator implementation.
Eight \gls{rfadc} channels are processed simultaneously at a clock rate of \ensuremath{{512}\,\mathrm{MHz}}, corresponding to eight samples per channel per cycle for a sample rate of \ensuremath{{4}\,\mathrm{GS/s}} per channel.
Data signals are represented with thick, black arrows, while the control inputs use thin, blue arrows.
The data flows from left (input) to right (output) as directed by the arrows.
Reused submodules, such as the timestamp generator, timer delay, and pipeline delay are highlighted with gray boxes.
The modules $<$low (all-below-low) and $>$high (any-above-high) compare the input data against low and high thresholds to decide when an event of interest has happened and when it is time to stop saving data after that event.
These modules provide the start and stop inputs to the finite state machine (FSM) that is described in Figure \ref{fig:disc}c.}
\label{fig:discdataflow}
\end{figure}

\subsubsection{Zero suppression with realtime sample discriminator}
\label{sec:discfsm}
Similar to the ``qualified buffer'' of \cite{xie_entangled_2023}, we implemented a sample discriminator based on \cite{watkins_splendaq_2024} to remove irrelevant samples and only save voltage waveforms surrounding events of interest.
Figures \ref{fig:disc} and \ref{fig:discdataflow} illustrate the operating principle and block diagram of the sample discriminator.
In order to reconstruct the original waveform, timetags for each event of interest are saved.
Similar to \cite{watkins_splendaq_2024} --- albeit operating in real-time --- our discriminator performs hysteretic comparison against two thresholds, which is useful for ensuring capture of tightly-spaced or small pulses that may occur in a malfunctioning circuit.
Furthermore, our discriminator can share triggers between multiple channels as illustrated in Figure \ref{fig:disc}b which allows the sample discriminator to trigger capture of a raw signal on a processed copy of that same signal. 
The ability to preprocess signals (e.g. with a filter or more complex signal processing module) before they enter the sample discriminator, combined with the trigger sharing technique, can enable arbitrarily complex triggering schemes that combine multiple channels.\footnote{The current firmware version has a fixed-weight finite-impulse-response filter, so more complicated triggering schemes would likely require implementation of a custom module.}
For example, a matched filter or lightweight neural network\footnote{A neural network may be able to handle non-stationary noise better than a matched filter, with the disadvantage of increased implementation complexity.} could be implemented, the output of which could be a likelihood estimate of a particular event occurring over the time window of the filter/network.
Such an output could be fed directly into the level discriminator or through additional logic to track multiple events over time.
With the trigger sharing functionality, these likelihood estimates or other higher-order signals could be used to trigger the capture of raw \gls{rfadc} data, allowing the user to analyze the original data surrounding events of interest after a measurement occurred.

The sample discriminator is fed a stream of data and valid signals.
By comparing the input data against a high and low threshold and using the comparison result to modulate the valid signal on the output stream, the sample discriminator can save samples of interest and discard others.
Deletion of uninteresting samples by deassertion of the output valid signal is illustrated in Figure \ref{fig:disc}a and \ref{fig:disc}b by the gray-shaded regions in the data out waveform.
The output data is delayed relative to the input data by a runtime configurable amount ($\tau_{\rm pre}$) to allow the user to save samples before an event of interest occurs.
Another parameter $\tau_{\rm post}$ can be used to save samples after an event of interest has ended.
The input valid signal is modulated by a state machine that transitions based on events generated by comparison of the input data with a high and low threshold.
Figure \ref{fig:disc}c shows the transition diagram for the state machine.
Each state machine (one for each of the eight capture channels) is initialized to the DISABLED state and outputs a valid signal whenever it is not in the DISABLED state.
Transitions to PRECAPTURE, CAPTURE, and/or POSTCAPTURE occur when internal start and stop signals and their delayed variants (start$_{\rm d}$ and stop$_{\rm d}$) are generated by comparing the input data with the high and low threshold.
PRECAPTURE is active when the input data has crossed the high threshold, but the delayed input data has not.
CAPTURE is active when the delayed input data has crossed the high threshold, but has not yet gone below the low threshold.
POSTCAPTURE is active for exactly one clock cycle after the CAPTURE state exits, and is only used when the capture is initiated by a digital trigger.
When configured to trigger based on the input signal value with non-zero pre- and post- event delays $\tau_{\rm pre}$ and $\tau_{\rm post}$, the state machine transitions from DISABLED, to PRECAPTURE, to CAPTURE, then back to DISABLED. If the combined delay $\tau_{\rm total} = \tau_{\rm pre} + \tau_{\rm post}$ is zero, the PRECAPTURE state is bypassed.
Figure \ref{fig:disc}a shows an example of the various state transitions based on signals generated by the input data and delay elements.
For digital triggers, the stop event is unused, so the POSTCAPTURE state is used to ensure the correct number of samples are saved.
Trigger sharing is implemented with crossbar multiplexers that allow the start/stop signals from one \gls{rfadc} channel to control the state machine for a different channel (see Figure \ref{fig:discdataflow}b).

Because of the very high sample rate of the \gls{rfadc}, instead of comparing individual samples against the thresholds, batches of eight samples are processed each clock cycle ($f_{\rm clk} = \ensuremath{{512}\,\mathrm{MHz}}$).
Batch-processing samples substantially simplifies the complexity of the module compared to an approach that processes samples individually.
Furthermore, batching in this way has little impact on the maximum deletion ratio for signals with very few events.
However, comparing multiple samples with the high and low thresholds can potentially introduce deadlock or race conditions if one does not select the right way to compare samples.
The operating principle of the threshold comparison works as follows: if any of the eight samples in a batch are greater than the high threshold --- ``any-above-high'' ---, then the sample discriminator exits the DISABLED state.
Similarly, if all eight samples are less than the low threshold --- ``all-below-low'' ---, then the sample discriminator returns to the DISABLED state.
The asymmetry in these event conditions (\textit{any}-above-high and \textit{all}-below-low) guarantees that they cannot both occur at the same time, although a simple fixed-priority arbiter could achieve the same effect if a symmetric scheme such as any-above-high and any-below-low were used.
The asymmetric condition was preferred over a symmetric scheme with a fixed-priority arbiter due the nature of the application: the penalty for accidentally deleting relevant data is considerably higher than the penalty for accidentally saving a few extra samples (e.g. 8) for each event.

As a user, one has several runtime-configurable parameters they can use to customize the operation of the sample discriminator. First, the user can specify how long data should be saved for before and after an event occurs ($\tau_{\rm pre}$, $\tau_{\rm post}$ in Figure \ref{fig:discdataflow}). Second, the user can configure which all-below-low and any-above-high events are passed to which state machine, allowing data values on one channel to control capture on another channel ("trigger source" control input in Figure \ref{fig:discdataflow}). Third, one can skip the level-based trigger and supply a digital trigger with an adjustable delay $\tau_{\rm digital}$ ("is digital" and $\tau_{\rm digital}$ inputs in Figure \ref{fig:discdataflow}). Finally, if desired, the user can completely bypass the discriminator ("bypass" input in Figure \ref{fig:discdataflow}).

\subsubsection{Event delays: saving data before and after events of interest}
The majority of the complexity in the design comes from allowing the user to save a runtime-configurable number of samples before the start (any-above-high) event.
This feature was added because it could be helpful while debugging a circuit to see if any odd behavior (potentially on other capture channels) occurs immediately before the input waveform crosses the high threshold.
As shown in Figure \ref{fig:discdataflow}, variable-delay pipelines and resettable counters are used to delay various data and trigger signals by a runtime-configurable amount (up to the pipeline depth or counter range set by its bit width).
By comparing the input stream with the high and low thresholds and operating on a delayed copy of the input stream, the discriminator can effectively look back in time to save samples before the start event.
The delays for the start and stop signals are generated differently, which affects the response of the circuit to multiple start or stop signals arriving within a short time interval.
The start signal uses a timer delay, which will only delay the final start signal in a burst of multiple start signals that occur in rapid succession (see Figure \ref{fig:disc}a for an example of two start events in a short window).
In comparison, the stop signal uses a pipeline delay, which will pass all inputs through to its output, allowing the first stop signal in a burst to transition the state machine.
This distinction is important because a timer delay for the stop signal would not allow the state machine to reset, since input data below the low threshold would continuously generate new stop signals, resetting the timer. In addition, a pipeline delay for the start signal would allow the state machine to ignore all but the first event in a short burst of start events, potentially leading to missed pulses.

The sample discriminator saves samples from the data stream at arbitrary, data-dependent times, so timing information for each sample needs to be known to reconstruct the original input signal.
Fortunately, the timing information for the majority of samples can be inferred, so a limited amount of extra storage is required for this information.
This information is provided in the form of a timestamp, generated from two counters: a monotonic timekeeping counter that increments every clock cycle (default 50-bit wide), and a sample index counter that increments for every valid output sample (default 14-bit wide).
The sample index counter is reset when a new data capture is initiated.
A timestamp is only output when the state machine transitions out of the DISABLED state, and later samples are assumed to immediately follow that timestamped sample, until the next timestamp is output.
Because a timestamp is not provided for every sample, the sample index counter is needed to determine which sample corresponds to the timestamp that was output.
The bit-widths of the two counters are automatically generated based on the depth of the capture memory.

\section{Analog frontend characterization}
\label{sec:afe}

\begin{figure}[htbp]
\centering
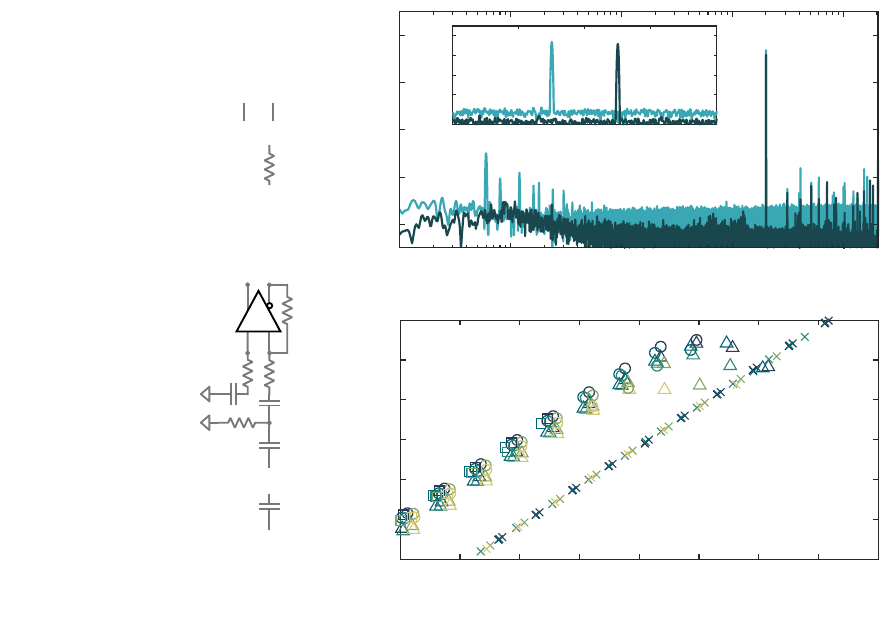
\caption{
Characterization of analog frontend shows reduced noise and anharmonic spur content compared with passive breakout board.
\textbf{(a)} Electrical schematic of analog frontend amplifier chains that perform conversion between single-ended and differential signals.
The transmit and receive paths are connected in loopback through various attenuators (\ensuremath{{-20}\,\mathrm{dB}}, \ensuremath{{-32}\,\mathrm{dB}}, and \ensuremath{{-57}\,\mathrm{dB}}).
Testing is performed by injecting a high-purity sinusoid generated with \gls{dds} and capturing the \gls{rfadc} output with a \gls{dma} buffer.
\textbf{(b)} Example spectrum for \ensuremath{{200}\,\mathrm{MHz}} excitation comparing the performance of the passive balun breakout board (XM500) to the active analog frontend (AFE).
The noise floor and some of the anharmonic spurs are reduced by the amplifiers.
The spectra are shifted by \ensuremath{{1}\,\mathrm{MHz}} in the inset for clarity.
\textbf{(c)}
The use of the low-noise amplifier in the active frontend improves \acrfull{sinad} over the passive balun by roughly \ensuremath{{20}\,\mathrm{dB}} with the \ensuremath{{0}\,\mathrm{dB}} floor limited by the \ensuremath{{3}\,\mathrm{dB}} noise figure of the low-noise amplifier.
}
\label{fig:afe_perf}
\end{figure}

Analog frontend (AFE) characterization was performed with single-tone loopback tests with the \gls{dds} signal generator and a \gls{dma} buffer to capture the data from the \gls{rfadc}.
The goal of this measurement was to quantify the amount of noise and nonlinearity introduced by the receive amplifiers, assuming that the \gls{rfdac} and transmit amplifiers can produce a sufficiently clean tone.
Figure \ref{fig:afe_perf}a shows an electrical diagram of the analog frontend in a loopback configuration with various external attenuators.
The transmit circuit is comprised of two fully-differential amplifiers which convert the \gls{rfdac} differential output to single ended without loading its internal biasing circuit.
The receive circuit is comprised of a low-noise amplifier, whose output is converted to a differential signal and passed through a \gls{vga} and anti-alias filter before being supplied to the \gls{rfadc}.
The AFE is compared with a passive breakout board (XM500) with \ensuremath{{10}\,\mathrm{MHz}}--\ensuremath{{1}\,\mathrm{GHz}} baluns for single-ended to differential conversion.
The receive-path amplifier chain reduces the noise floor and anharmonic spurs (i.e. frequency content that is not an integer multiple of the \ensuremath{{200}\,\mathrm{MHz}} excitation tone) compared to passive XM500 breakout board as shown in Figure \ref{fig:afe_perf}b.
Figure \ref{fig:afe_perf}c shows the \gls{sinad} ratio for sinusoidal tones at various frequencies and power levels.

The introduction of the amplifiers in the receive signal path improves \gls{sinad} by \ensuremath{{20}\,\mathrm{dB}} for small input signals, pushing it close to the thermal noise floor of \ensuremath{{-81}\,\mathrm{dBm}} for \ensuremath{{2}\,\mathrm{GHz}} bandwidth as shown in Figure \ref{fig:afe_perf}c.
The measured noise floor of approximately \ensuremath{{-79}\,\mathrm{dBm}} is consistent with the specified \ensuremath{{3}\,\mathrm{dB}} noise figure of the TRF37D73 amplifier.
Note that the linearity is limited (as illustrated by the saturation of \gls{sinad} for input power levels above \ensuremath{{-40}\,\mathrm{dBm}}) because the measurement was performed with a fixed setting for the LMH6401 \gls{vga}, so at higher input levels the \gls{rfadc} input saturated.
The TRF37D73 amplifier has an input compression power (P1dB$_{\rm in}$) of \ensuremath{{-4}\,\mathrm{dBm}}, so it should be possible to preserve linearity over more than \ensuremath{{70}\,\mathrm{dB}} of dynamic range above the noise floor by adjusting the VGA according to the input power.

As shown in Figure \ref{fig:afe_perf}a, the transmit path amplifier contains two stages to step down the common-mode voltage of the \gls{rfdac} without loading the on-chip \gls{rfdac} biasing network (which would degrade its frequency performance).
A fully differential amplifier was chosen due to its very high maximum stable bandwidth.
However, the single-ended to differential conversion circuit used in this portion of the circuit is prone to long-term drift, requiring recalibration of the offset by a few mV every few hours.
A better approach would be to use a lower bandwidth low-drift operational amplifier (opamp).
If additional high-frequency performance is desired, a diplexer could be used to combine the opamp output with a second AC-coupled path that uses a single-ended amplifier.
A diplexer would enable the use of an opamp with better offset/drift characteristics as well by reducing the opamp's required bandwidth. Regardless of potential improvements to the design of the analog frontend, its bandwidth, dynamic range, and noise performance are adequate for high-speed testing of devices that produce very small (e.g. sub-millivolt) output signals.

\section{Preliminary nanowire circuit tests}

\begin{figure}[htbp]
\centering
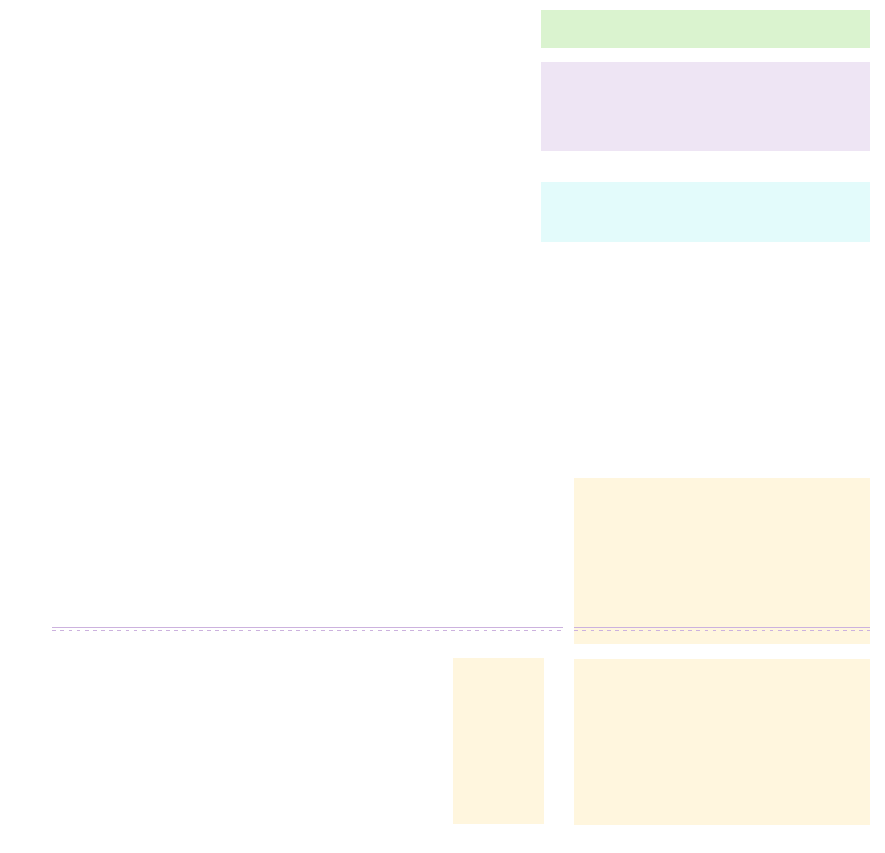
\caption{
Measurement of superconducting binary shift register.
\textbf{(a)} Electron micrographs of a nanocryotron (left) and the shift register (right).
\textbf{(b)} Electrical diagram of experimental setup. The \gls{rfsoc} \gls{fpga} and AFE are connected to the sample with an RF probe \cite{butters_digital_2022} cooled with liquid helium (see Figure \ref{fig:arch}a for a photo of the experimental setup).
\textbf{(c)} Waveform data comparing a \gls{cots} digitizer with a sample rate of \ensuremath{{500}\,\mathrm{MS/s}} \textbf{(i)} to the custom \gls{rfsoc} digitizer operating at \ensuremath{{4}\,\mathrm{GS/s}} \textbf{(ii,iii)}.
The time axis is shared within each column of subplots.
Each subplot includes the input stimulus: data and two-phase clock (top, gray) and digitized outputs: shunt$_1$, shunt$_2$ and out (bottom, colored).
The subplots on the right show \ensuremath{{50}\,\mathrm{ns}}-long subsections of the left subplots to illustrate the difference in sample rates on capture fidelity.
The digitized signal amplitude is given in mV, referred to the input of the analog frontend by dividing the digitized ADC value by the gain of the analog frontend.
When switching from the \gls{cots} to the \gls{rfsoc} digitizer, the sub-\ensuremath{{200}\,\mathrm{ps}} rise time of the output signal became visible, and the maximum clock rate the shift register could operate at was increased from \ensuremath{{83}\,\mathrm{MHz}} to \ensuremath{{200}\,\mathrm{MHz}} as shown in \textbf{(iii)}.
\textbf{(ii)} shows operation of the sample discriminator with $\tau_{\rm pre} = \tau_{\rm post} = 0$.
The high and low levels for the discriminator are indicated by solid and dashed horizontal lines.
}
\label{fig:experiment}
\end{figure}

To demonstrate the usefulness of the custom data acquisition system, it was used to test a previously-characterized binary shift register \cite{foster_superconducting_2023}.
The shift register is constructed from superconducting loops that store persistent currents to encode digital states.
Nanocryotrons act as switches to transfer the state between adjacent loops upon the application of a two-phase clock.\footnote{
The clock signal is a train of pulses, which can be generated with dedicated function or pulse generators.
However, characterization of some superconducting nanowire circuits requires generation of more complex waveforms \cite{butters_digital_2022} that can only be generated with an arbitrary waveform generator.
}
When there is a circulating current present in one of the loops, the sum of the circulating current and applied clock pulse exceed the critical current density in the nanocryotron channel constriction.
This causes the nanocryotron to become resistive, pushing the applied clock pulse into the next loop.
The resistive channel causes a voltage to develop, which can be measured to deduce the transition of state within the circuit (since directly measuring the circulating current would be impractical). Figure \ref{fig:experiment}a shows scanning electron micrographs of the shift register and a nanocryotron.

Figure \ref{fig:experiment}b shows an electrical diagram of the experimental setup used to evaluate the performance of the data acquisition system on a shift register with two superconducting loops.
The shift register chip was glued and wirebonded to a \gls{pcb}.
Surface-mount resistors were used to approximate current sources for the clocks and input and were soldered to the \gls{pcb} which was cooled to \ensuremath{{4.2}\,\mathrm{K}} using an RF probe \cite{butters_digital_2022} loaded in a liquid helium dewar.
The voltages across the output nanocryotron (out), shift nanocryotron (shunt$_2$), and input nanocryotron (shunt$_1$) were monitored.
For measurements with the \gls{cots} system from Keysight, two external LNA-2500 amplifiers from RF-Bay (\ensuremath{{\approx50}\,\mathrm{dB}} gain total) were used to amplify each output of the shift register (six LNAs required in total) before digitization.

The results obtained from the \gls{cots} digitizer and the custom digitizer presented in this work are compared in Figure \ref{fig:experiment}c.
The difference in sample rates is obvious between the two cases, with the $8\times$ higher sample rate and $10\times$ higher analog bandwidth of the \gls{rfsoc} system resolving the sub-\ensuremath{{200}\,\mathrm{ps}} rise time of the output nanocryotron.
Another phenomenon that becomes observable
at higher sample rates is the differentiation of the fast clock pulses, which occurs because of excess inductance in series with the nanocryotrons that forces the rapidly-changing current of the clock edge into the shunt resistor. This undesirable effect makes it more difficult to determine when the nanocryotron actually switched, and is only exacerbated at higher clock rates.
The effect is most clearly visible in the \ensuremath{{200}\,\mathrm{MHz}} test with the \gls{rfsoc} digitizer \ref{fig:experiment}c(iii), where the clock signal uses a \ensuremath{{100}\,\mathrm{ps}} rising and falling edge.
The fast edge results in substantial voltage pulses even when the nanocryotron does not switch.
Most importantly, however, is the nearly three-fold increase in maximum operating speed of the shift register: with the \gls{cots} equipment, the maximum clock rate attainable in experiment was \ensuremath{{83}\,\mathrm{MHz}} due to sample rate limitations of the \gls{awg}, but when using the data acquisition system presented in this work, the same circuit could operate at \ensuremath{{200}\,\mathrm{MHz}}, which was the highest frequency at which we observed it operate. At higher clock rates, even for very short clock/input pulses, the shift register would produce incorrect outputs, either switching when it shouldn't (i.e. 0$\rightarrow$1 errors) or not switching when it should (i.e. 1$\rightarrow$0 errors).
This is not a fundamental limit of nanocryotron devices, which have been demonstrated to operate at speeds exceeding \ensuremath{{600}\,\mathrm{MHz}} \cite{zheng_characterize_2019} (and individual nanowires have demonstrated speeds of \ensuremath{{2}\,\mathrm{GHz}} \cite{pearlman_gigahertz_2005}).
The limited operating frequency of the shift register underscores the importance of high-speed test equipment when designing circuits with novel devices: due to the way this particular circuit was designed, it is unable to operate at the maximum speed of the individual devices it is comprised of.
It should be noted that in addition to allowing us to probe the speed limits of this shift register circuit, the \gls{cots} digitizer system (\gls{awg}, digitizer, controller, chassis) as configured cost about seven times more than the ZCU111 \gls{rfsoc} and custom analog frontend (\$110,000 vs. \$16,000).

The operation of the sample discriminator is shown in Figure \ref{fig:experiment}c(iii).
Horizontal solid and dashed lines indicate the high (start) and low (stop) thresholds for controlling capture of data.
The pre- and post-event delays ($\tau_{\rm pre}$ and $\tau_{\rm post}$) are both set to zero.
However, a few samples before the signal crosses the upper threshold (or after the signal crosses the lower threshold) are visible.
This is because the sample discriminator processes batches of eight samples at a time at a clock rate of \ensuremath{{512}\,\mathrm{MHz}} to handle the high sample rate, so a few extra samples (at most 14 for each voltage pulse) that do not meet the threshold criteria may get saved.

\section{Conclusion and outlook}

This work presents a cost-effective, high-data-rate time-domain data acquisition system based on an \gls{rfsoc} with custom firmware for timetagging events of interest and sparse data capture.
The data acqusition system will enable more sophisticated testing of superconducting circuits, ushering in the development of more advanced circuits for performing complex tasks.

Although designed with testing of nanocryotron circuits in mind, the performance and features of this data acquisition system have several other useful applications.
For one, the low noise analog frontend is capable of testing \gls{jj}-based circuits at gigahertz clock speeds, provided the circuit output is made suitably large (e.g. \ensuremath{{100}\,\upmu\mathrm{V}}) with Suzuki stacks \cite{suzuki_josephson_1988} and/or cryogenic amplifiers.
This opens up the possibility of using it to test \gls{rsfq} and \gls{aqfp} logic circuits.
Although \glspl{jj} are more mature than superconducting nanowires, the ability to simultaneously control and monitor many inputs and outputs could be very useful, particularly in situations where specialized high-speed CMOS testing equipment like the bit-error-rate tester used in \cite{ayala_mana_2021} make it difficult to fully evaluate the possible input space of a circuit.
In addition to electronic circuits, the timetagging acquisition aspect of this work could be useful for detector readout.
Indeed, a similar scheme is already used by \cite{xie_entangled_2023}.
However, the applicability of this work to frequency-multiplexed readout is fairly limited, since the firmware would need to be heavily modified to introduce polyphase filter banks to perform demodulation of frequency-multiplexed resonator signals \cite{smith_high-throughput_2021}, and one would need to adjust the parameter configuring the number of channels for the sample discriminator to match the number of resonators.
For this reason, existing specialized \gls{fpga}-based solutions for readout of frequency-multiplexed superconducting detectors would be more appropriate for these applications \cite{smith_mkidgen3_2024, liu_development_2024, yan_readout_2024}.
However, the approach of saving the raw signal from the detector using the trigger-sharing mechanism presented in this work may still be useful for detector analysis, as it allows for recording of a minimal amount of data for subsequent postprocessing that would be impractical to run in real time.
For \glspl{snspd} used as particle detectors \cite{polakovic_unconventional_2020}, this may provide insight into what caused the detector to fire in the first place, because the pulse shape of the detection waveform varies depending on the type of particle absorbed \cite{armstrongJlab}.
It could also be useful for mitigating time-walk effects at very high count rates in \glspl{snspd} or other current-biased detectors or transducers \cite{mueller_timewalk_2023}.
In addition, the high sensitivity of the analog frontend could enable the data acquisition system to discriminate between single- and multi-photon events for an impedance-matched \gls{snspd} \cite{zhu_resolving_2020}.

The high sample rate of the dataconverters in the \gls{rfsoc} is fast enough to probe the speed limits of technologies that are dominated by thermal processes, such as nanocryotrons.
As a result, the tool can be used to measure the physical processes that occur in these devices, which is helpful for developing more accurate simulation models and deepening our understanding of them.
Experiments like this have been done previously, with high-sample rate \gls{awg}s and oscilloscopes \cite{zheng_characterize_2019}.
However the ability to generate much shorter pulses with the multi-GS/s \glspl{rfdac} on the \gls{rfsoc} platform would enable accurate sub-threshold measurements where multiple pulses of current together switch a nanocryotron, but a single pulse does not.

The firmware was intentionally developed to be extensible, allowing new features to be added to further improve the testing capabilities of the data acquisition system.
Even though a ZCU111 \gls{rfsoc} was used, virtually all of the firmware is written to be device-agnostic, and could be used in other \gls{fpga}s, with the only significant aspect requiring modification being the dataconverter (\gls{rfadc}/\gls{rfdac}) interface.
It should be noted that for a smaller number of channels, products like the RFSoC4x2 from RealDigital may be a more cost-effective option if one does not require as many \glspl{rfdac} or \gls{rfdac} operation down to DC.
In the future, we plan to use the data acquisition system presented in this work to study the error rate of more complex superconducting circuits over various operating conditions such as temperature and external magnetic field.

\appendix
\printnoidxglossary[type=\acronymtype,style=index,nogroupskip]{}

\paragraph{Acknowledgments}
The authors thank Camron Blackburn, Malick Sere, and Phillip Donnie Keathley for helpful comments and feedback during the preparation of this manuscript. The authors thank Stephen Nagle of MIT TJ Rogers Laboratory for his assistance and guidance with assembly of the analog frontend. This work was funded by the DOE Office of Science Research Program for Microelectronics Codesign through the project ``Hybrid Cryogenic Detector Architectures for Sensing and Edge Computing enabled by new Fabrication Processes'' (LAB 21-2491). RF acknowledges funding from the Alan McWhorter fellowship. OM acknowledges funding from NDSEG fellowship. AS acknowledges funding from the NSF GRFP.

\paragraph{Disclaimer}
This report was prepared as an account of work sponsored by an agency of the United States Government. Neither the United States Government nor any agency thereof, nor any of their employees, makes any warranty, express or implied, or assumes any legal liability or responsibility for the accuracy, completeness, or usefulness of any information, apparatus, product, or process disclosed, or represents that its use would not infringe privately owned rights. Reference herein to any specific commercial product, process, or service by trade name, trademark, manufacturer, or otherwise does not necessarily constitute or imply its endorsement, recommendation, or favoring by the United States Government or any agency thereof. The views and opinions of authors expressed herein do not necessarily state or reflect those of the United States Government or any agency thereof.



\bibliographystyle{JHEP}
\bibliography{biblio.bib}

\providecommand{\href}[2]{#2}\begingroup\raggedright\begin{thebibliography}{10}

\bibitem{wachter_microprocessor_2017}
S.~Wachter, D.K.~Polyushkin, O.~Bethge and T.~Mueller, \emph{A microprocessor
  based on a two-dimensional semiconductor},
  \href{https://doi.org/10.1038/ncomms14948}{\emph{Nature Communications}
  {\bfseries 8} 14948}.

\bibitem{peng_carbon_2019}
L.-M.~Peng, Z.~Zhang and C.~Qiu, \emph{Carbon nanotube digital electronics},
  \href{https://doi.org/10.1038/s41928-019-0330-2}{\emph{Nature Electronics}
  {\bfseries 2} 499}.

\bibitem{polonsky_new_1993}
S.~Polonsky, V.~Semenov, P.~Bunyk, A.~Kirichenko, A.~Kidiyarov-Shevchenko,
  O.~Mukhanov et~al., \emph{New {RSFQ} circuits (josephson junction digital
  devices)}, \href{https://doi.org/10.1109/77.233530}{\emph{{IEEE} Transactions
  on Applied Superconductivity} {\bfseries 3} 2566}.

\bibitem{ayala_mana_2021}
C.L.~Ayala, T.~Tanaka, R.~Saito, M.~Nozoe, N.~Takeuchi and N.~Yoshikawa,
  \emph{{MANA}: A monolithic adiabatic {iNtegration} architecture
  microprocessor using 1.4-{zJ}/op unshunted superconductor josephson junction
  devices}, \href{https://doi.org/10.1109/JSSC.2020.3041338}{\emph{{IEEE}
  Journal of Solid-State Circuits} {\bfseries 56} 1152}.

\bibitem{li_analogue_2018}
C.~Li, M.~Hu, Y.~Li, H.~Jiang, N.~Ge, E.~Montgomery et~al., \emph{Analogue
  signal and image processing with large memristor crossbars},
  \href{https://doi.org/10.1038/s41928-017-0002-z}{\emph{Nature Electronics}
  {\bfseries 1} 52}.

\bibitem{koshelets_experimental_1987}
V.~Koshelets, K.~Likharev, V.~Migulin, O.~Mukhanov, G.~Ovsyannikov, V.~Semenov
  et~al., \emph{Experimental realization of a resistive single flux quantum
  logic circuit}, \href{https://doi.org/10.1109/TMAG.1987.1064953}{\emph{{IEEE}
  Transactions on Magnetics} {\bfseries 23} 755}.

\bibitem{kaplunenko_experimental_1989}
V.~Kaplunenko, M.~Khabipov, V.~Koshelets, K.K.~Likharev, O.~Mukhanov and
  V.~Semenov, \emph{Experimental study of the {RSFQ} logic elements},
  \href{https://doi.org/10.1109/20.92422}{\emph{{IEEE} Transactions on
  Magnetics} {\bfseries 25} 861}.

\bibitem{mccaughan_superconducting-nanowire_2014}
A.N.~{McCaughan} and K.K.~Berggren, \emph{A superconducting-nanowire
  three-terminal electrothermal device},
  \href{https://doi.org/10.1021/nl502629x}{\emph{Nano Letters} {\bfseries 14}
  (2014) 5748}.

\bibitem{steinhauer_progress_2021}
S.~Steinhauer, S.~Gyger and V.~Zwiller, \emph{Progress on large-scale
  superconducting nanowire single-photon detectors},
  \href{https://doi.org/10.1063/5.0044057}{\emph{Applied Physics Letters}
  {\bfseries 118} (2021) 100501}.

\bibitem{polakovic_unconventional_2020}
T.~Polakovic, W.~Armstrong, G.~Karapetrov, Z.-E.~Meziani and V.~Novosad,
  \emph{Unconventional applications of superconducting nanowire single photon
  detectors}, \href{https://doi.org/10.3390/nano10061198}{\emph{Nanomaterials}
  {\bfseries 10} (2020) 1198}.

\bibitem{huang_monolithic_2024}
Y.-H.~Huang, Q.-Y.~Zhao, H.~Hao, N.-T.~Liu, Z.~Liu, J.~Deng et~al.,
  \emph{Monolithic integrated superconducting nanowire digital encoder},
  \href{https://doi.org/10.1063/5.0202827}{\emph{Applied Physics Letters}
  {\bfseries 124} (2024) 192601}.

\bibitem{butters_scalable_2021}
B.A.~Butters, R.~Baghdadi, M.~Onen, E.A.~Toomey, O.~Medeiros and K.K.~Berggren,
  \emph{A scalable superconducting nanowire memory cell and preliminary array
  test}, \href{https://doi.org/10.1088/1361-6668/abd14e}{\emph{Supercond. Sci.
  Technol.} {\bfseries 34} (2021) 035003}.

\bibitem{foster_superconducting_2023}
R.A.~Foster, M.~Castellani, A.~Buzzi, O.~Medeiros, M.~Colangelo and
  K.K.~Berggren, \emph{A superconducting nanowire binary shift register},
  \href{https://doi.org/10.1063/5.0144685}{\emph{Applied Physics Letters}
  {\bfseries 122} (2023) 152601}.

\bibitem{buzzi_nanocryotron_2023}
A.~Buzzi, M.~Castellani, R.A.~Foster, O.~Medeiros, M.~Colangelo and
  K.K.~Berggren, \emph{A nanocryotron memory and logic family},
  \href{https://doi.org/10.1063/5.0144686}{\emph{Applied Physics Letters}
  {\bfseries 122} (2023) 142601}.

\bibitem{castellani_nanocryotron_2024}
M.~Castellani, O.~Medeiros, R.A.~Foster, A.~Buzzi, M.~Colangelo, J.C.~Bienfang
  et~al., \emph{Nanocryotron ripple counter integrated with a superconducting
  nanowire single-photon detector for megapixel arrays},
  \href{https://doi.org/10.1103/PhysRevApplied.22.024020}{\emph{Physical Review
  Applied} {\bfseries 22} (2024) 024020}.

\bibitem{zhao_compact_2018}
Q.-Y.~Zhao, E.A.~Toomey, B.A.~Butters, A.N.~{McCaughan}, A.E.~Dane, S.-W.~Nam
  et~al., \emph{A compact superconducting nanowire memory element operated by
  nanowire cryotrons},
  \href{https://doi.org/10.1088/1361-6668/aaa820}{\emph{Superconductor Science
  and Technology} {\bfseries 31} (2018) 035009}.

\bibitem{zheng_superconducting_2020}
K.~Zheng, Q.-Y.~Zhao, H.-Y.-B.~Lu, L.-D.~Kong, S.~Chen, H.~Hao et~al., \emph{A
  superconducting binary encoder with multigate nanowire cryotrons},
  \href{https://doi.org/10.1021/acs.nanolett.0c00498}{\emph{Nano Letters}
  {\bfseries 20} (2020) 3553}.

\bibitem{toomey_superconducting_2020}
E.~Toomey, K.~Segall, M.~Castellani, M.~Colangelo, N.~Lynch and K.K.~Berggren,
  \emph{Superconducting nanowire spiking element for neural networks},
  \href{https://doi.org/10.1021/acs.nanolett.0c03057}{\emph{Nano Letters}
  {\bfseries 20} (2020) 8059}.

\bibitem{huang_splitter_2023}
Y.-H.~Huang, Q.-Y.~Zhao, S.~Chen, H.~Hao, H.~Wang, J.-W.~Guo et~al.,
  \emph{Splitter trees of superconducting nanowire cryotrons for large
  fan-out}, \href{https://doi.org/10.1063/5.0139791}{\emph{Applied Physics
  Letters} {\bfseries 122} (2023) 092601}.

\bibitem{wang_attojoule_2025}
H.~Wang, N.~Noordzij, M.~Mikhailov, S.~Steinhauer, T.~Descamps, E.~Oksenberg
  et~al., \emph{Attojoule superconducting thermal logic and memories},
  \href{https://doi.org/10.1021/acs.nanolett.4c06545}{\emph{Nano Letters}
  {\bfseries 25} (2025) 4401}.

\bibitem{castellani_superconducting_2025}
M.~Castellani, O.~Medeiros, A.~Buzzi, R.A.~Foster, M.~Colangelo and
  K.K.~Berggren, \emph{A superconducting full-wave bridge rectifier},
  \href{https://doi.org/10.1038/s41928-025-01376-4}{\emph{Nature Electronics}
  {\bfseries 8} (2025) 417}.

\bibitem{keysight_pxi_nodate}
Keysight, ``{PXI} and {AXIe} products and solutions catalog - {September}
  2020.''

\bibitem{national_instruments_ni_nodate}
{National Instruments}, ``{NI} {PXI} product catalog.''

\bibitem{zinoviev_octopux_1997}
D.~Zinoviev and Y.~Polyakov, \emph{Octopux: an advanced automated setup for
  testing superconductor circuits},
  \href{https://doi.org/10.1109/77.622039}{\emph{{IEEE} Transactions on Applied
  Superconductivity} {\bfseries 7} (1997) 3240}.

\bibitem{butters_digital_2022}
B.A.~Butters, \emph{Digital and Microwave Superconducting Electronics and
  Experimental Apparatus}, {PhD} thesis, Massachusetts Institute of Technology,
  2022.

\bibitem{stefanazzi_qick_2022}
L.~Stefanazzi, K.~Treptow, N.~Wilcer, C.~Stoughton, C.~Bradford, S.~Uemura
  et~al., \emph{The {QICK} (quantum instrumentation control kit): Readout and
  control for qubits and detectors},
  \href{https://doi.org/10.1063/5.0076249}{\emph{Review of Scientific
  Instruments} {\bfseries 93} (2022) 044709}.

\bibitem{smith_mkidgen3_2024}
J.P.~Smith, J.I.~Bailey, {III}, A.~Cuda, N.~Zobrist and B.A.~Mazin,
  \emph{{MKIDGen}3: Energy-resolving, single-photon-counting microwave kinetic
  inductance detector readout on a radio frequency system-on-chip},
  \href{https://doi.org/10.1063/5.0225768}{\emph{Review of Scientific
  Instruments} {\bfseries 95} (2024) 114705}.

\bibitem{liu_development_2024}
C.~Liu, Z.~Ahmed, S.W.~Henderson, R.~Herbst, L.~Ruckman and T.~Satterthwaite,
  \emph{Development of {RFSoC}-based direct sampling highly multiplexed
  microwave {SQUID} readout for {CMB} and submillimeter surveys},  in
  \emph{Millimeter, {Submillimeter}, and {Far}-{Infrared} {Detectors} and
  {Instrumentation} for {Astronomy} {XII}}, vol.~13102, pp.~422--429, SPIE,
  Aug., 2024, \href{https://doi.org/10.1117/12.3019317}{DOI}.

\bibitem{kerman_kinetic_inductance_2006}
A.J.~Kerman, E.A.~Dauler, W.E.~Keicher, J.K.W.~Yang, K.K.~Berggren,
  G.~Gol’tsman et~al., \emph{Kinetic-inductance-limited reset time of
  superconducting nanowire photon counters},
  \href{https://doi.org/10.1063/1.2183810}{\emph{Applied Physics Letters}
  {\bfseries 88} (2006) 111116}.

\bibitem{xie_entangled_2023}
S.~Xie, L.~Stefanazzi, C.~Wang, C.~Peña, R.~Valivarthi, L.~Narváez et~al.,
  \emph{Entangled photon pair source demonstrator using the quantum
  instrumentation control kit system},
  \href{https://doi.org/10.1109/JQE.2023.3302926}{\emph{{IEEE} Journal of
  Quantum Electronics} {\bfseries 59} (2023) 1}.

\bibitem{pg269}
{AMD/Xilinx}, ``{PG269} {Zynq} {UltraScale+} {RFSoC} {RF} data converter v2.6
  gen 1/2/3/{DFE} {LogiCORE} {IP} product guide.''

\bibitem{foster_rfsoc_daq}
R.~Foster, ``{rfsoc\_daq}.''
\newblock https://doi.org/10.5281/zenodo.15264422.

\bibitem{noauthor_pynq_nodate}
{AMD/Xilinx}, ``{https://www.pynq.io}.''

\bibitem{crockett_exploring_2019}
L.H.~Crockett, D.~Northcote, C.~Ramsay, F.D.~Robinson and R.W.~Stewart,
  \emph{Exploring Zynq {MPSoC}: with {PYNQ} and machine learning applications},
  Strathclyde Academic Media (2019).

\bibitem{abbas_latency_2018}
M.~Abbas and V.~Betz, \emph{Latency insensitive design styles for {FPGAs}},  in
  \emph{2018 28th International Conference on Field Programmable Logic and
  Applications ({FPL})}, pp.~360--3607, 2018,
  \href{https://doi.org/10.1109/FPL.2018.00068}{DOI}.

\bibitem{mehta_xilinx_nodate}
N.~Mehta, ``Xilinx {FPGA} embedded memory advantages.''

\bibitem{wp477}
{Xilinx}, ``{WP477} {UltraRAM}: Breakthrough embedded memory integration on
  {UltraScale+} devices.''

\bibitem{foster_scaling_2023}
R.A.~Foster, \emph{Scaling of nanocryotron superconducting logic},  {MEng}
  thesis, Massachusetts Institute of Technology, 2023.

\bibitem{watkins_splendaq_2024}
S.L.~Watkins, \emph{{SPLENDAQ}: A detector-agnostic data acquisition system for
  small-scale physics experiments},
  \href{https://doi.org/10.1007/s10909-023-03021-w}{\emph{Journal of Low
  Temperature Physics} {\bfseries 214} (2024) 133}.

\bibitem{zheng_characterize_2019}
K.~Zheng, Q.-Y.~Zhao, L.-D.~Kong, S.~Chen, H.-Y.-B.~Lu, X.-C.~Tu et~al.,
  \emph{Characterize the switching performance of a superconducting nanowire
  cryotron for reading superconducting nanowire single photon detectors},
  \href{https://doi.org/10.1038/s41598-019-52874-3}{\emph{Scientific Reports}
  {\bfseries 9} (2019) 16345}.

\bibitem{pearlman_gigahertz_2005}
A.~Pearlman, A.~Cross, W.~Slysz, J.~Zhang, A.~Verevkin, M.~Currie et~al.,
  \emph{Gigahertz counting rates of {NbN} single-photon detectors for quantum
  communications}, \href{https://doi.org/10.1109/TASC.2005.849926}{\emph{{IEEE}
  Transactions on Applied Superconductivity} {\bfseries 15} (2005) 579}.

\bibitem{suzuki_josephson_1988}
H.~Suzuki, A.~Inoue, T.~Imamura and S.~Hasuo, \emph{A josephson driver to
  interface josephson junctions to semiconductor transistors},  in
  \emph{Technical Digest., International Electron Devices Meeting},
  pp.~290--293, 1988, \href{https://doi.org/10.1109/IEDM.1988.32814}{DOI}.

\bibitem{smith_high-throughput_2021}
J.P.~Smith, J.I.~Bailey, J.~Tuthill, L.~Stefanazzi, G.~Cancelo, K.~Treptow
  et~al., \emph{A high-throughput oversampled polyphase filter bank using
  vivado {HLS} and {PYNQ} on a {RFSoC}},
  \href{https://doi.org/10.1109/OJCAS.2020.3041208}{\emph{{IEEE} Open Journal
  of Circuits and Systems} {\bfseries 2} (2021) 241}.

\bibitem{yan_readout_2024}
X.~Yan, F.~Liu, R.~Duan, X.~Ma, R.~Fan, X.~Wu et~al., \emph{Readout system for
  frequency-division multiplexing superconducting detector arrays},
  \href{https://doi.org/10.1007/s10909-024-03153-7}{\emph{Journal of Low
  Temperature Physics} {\bfseries 216} (2024) 626}.

\bibitem{armstrongJlab}
W.~Armstrong, ``Superconducting nanowire particle detectors.'' Jefferson
  Laboratory: Applications of Superconducting Electronics and Detectors
  Workshop, 11, 2022.

\bibitem{mueller_timewalk_2023}
A.~Mueller, E.E.~Wollman, B.~Korzh, A.D.~Beyer, L.~Narvaez, R.~Rogalin et~al.,
  \emph{Time-walk and jitter correction in {SNSPDs} at high count rates},
  \href{https://doi.org/10.1063/5.0129147}{\emph{Applied Physics Letters}
  {\bfseries 122} (2023) 044001}.

\bibitem{zhu_resolving_2020}
D.~Zhu, M.~Colangelo, C.~Chen, B.A.~Korzh, F.N.C.~Wong, M.D.~Shaw et~al.,
  \emph{Resolving photon numbers using a superconducting nanowire with
  impedance-matching taper},
  \href{https://doi.org/10.1021/acs.nanolett.0c00985}{\emph{Nano Letters}
  {\bfseries 20} 3858}.

\end{thebibliography}\endgroup

\end{document}